\documentstyle[preprint,prc,aps,epsf]{revtex}
\begin{document}
\include{def}
\def\v#1{\mbox{\boldmath$#1$}}
\newcommand{\smod}{******START MODIFICATION****** : }
\def\ket#1{|#1 \rangle}
\def\bra#1{\langle #1|}

\draft
\vspace{2cm}

\title{
Nonmesonic Weak Decay of $\Lambda$ Hypernuclei within a Nuclear
Matter Formalism}

\author{E. Bauer and F. Krmpoti\'c}

\address{
Departamento de F\'{\i}sica, C.C. 67, Facultad de Ciencias Exactas,\\
Universidad Nacional de La Plata,\\
La Plata, 1900, Argentina\\}

\maketitle

\date{\today}

\begin{abstract}
The nonmesonic weak decay of $\Lambda$ hypernuclei using
nonrelativistic nuclear matter is studied. As the basic building block we
use the Polarization Propagator Method	developed by Oset and Salcedo.
 It is shown that the exact calculation of exchange
terms is required. Using the Local Density Approximation we
evaluate the nonmesonic decay width for $^{12}_{\Lambda}C$ and
compare the result with a finite nucleus calculation, obtaining a
qualitative agreement.
\end{abstract}

\vspace{1.3cm}


\vspace{.5cm}


\newpage
\section{INTRODUCTION}
\label{INTR}

The $\Lambda$ decays within the hypernuclei via two
mechanisms: the first one is called mesonic decay, where
the product of the disintegration is a nucleon plus a pion.
The mesonic decay rate $\Gamma_M \equiv
\Gamma(\Lambda \rightarrow N \pi)$ is also the only
process in the $\Lambda$ free decay, $\Gamma^0$. The
second mechanism is the nonmesonic decay, where the
meson is absorbed by a nucleon. In this case the final
product are two nucleons, $\Gamma_{NM} \equiv
\Gamma(\Lambda N \rightarrow NN)$. Due to the Pauli
principle the mesonic decay is strongly blocked for $A
\geq 4$. At variance, this effect is not present in the
nonmesonic channel.  In principle, more than two
nucleons can emerge in the nonmesonic decay. In this
work we concentrate ourselves only on two nucleon emission.
In this case, the corresponding transition rates are
stimulated either by protons, $\Gamma_p \equiv
\Gamma(\Lambda p \rightarrow n p)$, or by neutrons,
$\Gamma_n \equiv \Gamma(\Lambda n \rightarrow n n)$.
The total nonmesonic decay rate is, $\Gamma_{NM} =
\Gamma_n + \Gamma_p $. While theory fairly accounts
for the experimental values for the total rate, the same is
not true for the ratio $\Gamma_{n/p} \equiv \Gamma_n /
\Gamma_p$, where theory underestimates data ($0.5 \leq
\Gamma^{exp}_{n/p} \leq 2$).

Several models have been
proposed to explain $\Gamma^{exp}_{n/p}$.
A good review of
the present status of the art can be found in Refs.~\cite{os98} and
\cite{al01}. Historically, Block and Dalitz (Refs.~\cite{da62},
\cite{bl63}) developed a phenomenological model (see
also Refs. \cite{do87}-\cite{al00}). From that point,
microscopic models have been explored. The first one is due to
Adams \cite{ad67}, who uses nuclear matter, one pion
exchange model (OPE), $\Delta I = 1/2$-$\Lambda N
\pi$ couplings and short range correlations (SRC). In the work
of Adams, the $\Lambda N \pi$ coupling was too
small to reproduce the $\Lambda$ free lifetime. This
mistake was corrected by McKellar and Gibson
\cite{mc84}, whom also included the $\rho$-meson. Also
using nuclear matter, Dubach {\it et al.} \cite{du86}
introduced a one meson exchange model (OME) with
$\pi$,	$\eta$,  $K$, $\rho$,  $\omega$ and $K^*$-mesons.

Oset and Salcedo \cite{os85} developed the polarization
propagator method (PPM). This scheme allows an unified
treatment of mesonic and nonmesonic channels. It is
performed in nuclear matter with the addition of the local
density approximation (LDA). In
the PPM one writes the expression for $\Lambda$
self-energy where, in principle, propagators for nucleons and mesons can
include correlations of all kinds. In Ref.
\cite{os85} propagators are dressed with correlations
of the Ramdom
Phase Approximation (RPA) type. The PPM is
further developed in Refs. \cite{al91}-\cite{al00c}.
In addition of the above-mentioned
mesons in Ref. \cite{ji01} two-pion correlations are also
considered. In a similar spirit of the PPM, Alberico and
Garbarino \cite{al01} employed the bosonic loop
expansion (BLE). These formalisms are particularly suitable
when more than two nucleons emerge from the
desintegration process.

Alternatively, some authors have employed finite nucleus wave
functions instead of plane waves. This method is usually called
Wave Function Method (WFM) (see Refs. \cite{ni93}-\cite{ba02}).
The method makes use of shell model nuclear and
hypernuclear wave function, as well as pion wave functions
generated by pion-nucleus optical potential. Finally, let us mention
that it was also included the quark degree of freedom to study this
problem \cite{sa00}, \cite{ch83}-\cite{in98}.

This list of works does not pretend to be complete. We have tried to
present the main physical ingredients considered in the
literature to deal with the $\Lambda$-weak decay. Even
though, the results for the ratio $\Gamma_{n/p}$ remains
unsatisfactory. Some of the above mentioned issues are still in its
preliminary stages and need further developments. Among which,
we can mentioned the two-nucleon stimulated process
(Refs. \cite{al91}-\cite{al00b}), the inclusion
of interaction terms that violates the isospin $\Delta I = 1/2$ rule
(Refs. \cite{go97}, \cite{pa98}, \cite{in98}),
and the quark degree of freedom.

To deal with many body correlations like the RPA ones,
a nuclear matter formalism is preferred over the
WFM due to the difficulties in dealing with continuous wave
functions within a WFM and also due to the big number of
configurations involve. For this reason we have worked out a nuclear
matter formalism. The first striking point is that most of the nuclear
matter works usually neglect exchange terms or evaluated them
in an approximate way.
Due to this, we have explored a nuclear matter scheme which contains exchange terms
and leads to results compatible with those of
finite nucleus.

The present work is organized as follows. In Sect. II, an overview of
the PPM is done and we present a nuclear matter formalism which includes
exchange terms. In Sect. III we show results for the nonmesonic
$\Lambda$ decay together with a comparison with others authors.
Finally, in Sect. IV some conclusions are drawn.

\section{FORMALISM}
\label{FORM}

We begin this section by briefly summarizing the main points of
the PPM formalism introduced by Oset and Salcedo \cite{os85}.
We first point out the advantages of the scheme, as well as
its limitations. Afterwards we describe the formalism used here.

The PPM gives an unified description of both the mesonic
and nonmesonic decay rates. Nonrelativistic nuclear
matter is employed. The basic idea of the PPM is to
evaluate the total decay rate $\Gamma_{\Lambda}=\Gamma_{M}+\Gamma_{NM}$,
using the imaginary part
of the $\Lambda$-self energy diagram of Fig. 1, as

\begin{equation}
\label{decw}
\Gamma_{\Lambda}(\v{k},k_F) = -2 Im \Sigma_{\Lambda}.
\end{equation}
where $\v{k}$ and $k_F$ are the $\Lambda$ and Fermi momentum, respectively.
The connection with $\Gamma_{\Lambda}$ is given by
\begin{equation}
\label{decwpar2}
\Gamma_{\Lambda}(k_F) = \int d \v{k} \,
\Gamma_{\Lambda}(\v{k}, k_F) \;  |\psi_{\Lambda}(\v{k})|^2 \,
\end{equation}
where for the $\Lambda$ wave function $\psi_{\Lambda}(\v{k})$, we
take the $1s_{1/2}$ wave function of a harmonic oscilator.
To evaluate $\Gamma_{\Lambda}$ for a particular nuclei one uses either an effective Fermi
momentum or the Local Density Approximation (LDA) \cite{os85}. In the last case
$k_F$ is  spatially dependent and the transition rate reads
\begin{equation}
\label{decwpar3}
\Gamma_{\Lambda} = \int d \v{r} \,
\Gamma_{\Lambda}( k_F(r)) \;  |\widetilde{\psi}_{\Lambda}(\v{r})|^2 \,
\end{equation}
where $\widetilde{\psi}_{\Lambda}(\v{r})$ is the Fourier transform of
$\psi_{\Lambda}(\v{k})$.

 The  $\Lambda N \pi$ vertex in Fig. 1 is described by the weak Hamiltonian
\begin{equation}
\label{Hamilt}
{\cal H}_{\Lambda N \pi} = i G_F m_{\pi}^2 \bar{\psi}_N (A_{\pi}
+B_{\pi} \gamma_5) \v{\tau} \cdot \v{\phi}_{\pi} \psi_{\Lambda} + h.c.
\end{equation}
where $ G_F m_{\pi}^2 = 2.21 \times 10^{-7}$ and the constants
$A_{\pi}=1.05$ and $B_{\pi}=-7.15$ are the parity violating and
parity conserving couplings constants \cite{go97}, respectively.
In Eq.~(\ref{Hamilt}) we assume the $\Delta I=1/2$ rule by taking
the hyperon as an isospin spurion with $I_3=-1/2$. The Hamiltonian
for the strong $NN \pi$ vertex, which will be used latter,  is given by
\begin{equation}
\label{Hamilts}
{\cal H}_{\pi NN} = i g_{NN \pi} \bar{\psi}_N \gamma_5 \v{\tau} \cdot \v{\phi}_{\pi}
\psi_{N}
\end{equation}
where the value of the strong-coupling constant is $g_{NN
\pi}=13.3$.

Using the standard Feynman rules, one gets in the
nonrelativistic limit,
\begin{equation}
\label{self1}
\Sigma_{\Lambda}(\v{k}, k_F) = 3 i (G_F m_{\pi}^2)^2
\int \frac{d^4 q}{(2 \pi)^4}\ (A_{\pi}^2 + \frac{B_{\pi}^2}{ 4
\bar{M}^2}\ \mbox{\boldmath $q$}^2 ) F_{\pi}^2 (q) G_N (k - q)
G_{\pi}(q),
\end{equation}
where the nucleon and pion propagators in the nuclear
medium are, respectively,

\begin{equation}
\label{nprop}
G_N(p) = \frac{\theta(|\mbox{\boldmath $p$}| - k_F)}
{p_0 - E_N(\mbox{\boldmath $p$})- V_N + i
\varepsilon} +
\frac{\theta(k_F - |\mbox{\boldmath $p$}| )}
{p_0 - E_N(\mbox{\boldmath $p$})- V_N - i
\varepsilon},
\end{equation}
and
\begin{equation}
\label{pprop}
G_{\pi}(q) = \frac{1}
{q_0^2 - \mbox{\boldmath $q$}^2 - m_{\pi}^2 -
\Sigma_{\pi}(q)}.
\end{equation}
Here, $p=(p_0, \mbox{\boldmath $p$})$ and $q=(q_0,
\mbox{\boldmath $q$})$ denote the energy-momentum
four-vector, $E_N$ is the nucleon total free energy,
$V_N$ is the nucleon binding energy and $\Sigma_{\pi}$
is the pion self-energy in nuclear matter.
The constant $\bar{M}$ is the average between the
nucleon and $\Lambda$ masses and
$k_F$ is the Fermi momentum.
Plane waves for nucleons were employed in the derivation
of Eq.~(\ref{self1}), together with a step
function which tells us if the nucleon is a particle or a
hole. The finite nucleon size is  shaped by the monopole form factor
\begin{equation}
\label{ffact}
F_{\pi}(q) = \frac{\Lambda_{\pi}^2 - m_{\pi}^2}
{\Lambda_{\pi}^2 - q_0^2 + \mbox{\boldmath $q$}^2},
\end{equation}
where $m_{\pi}$ is the pion mass and $\Lambda_{\pi}=1.3$ GeV.

The PPM is based on the behavior of the pion in the
nuclear medium.
The Feynman diagram displayed in  Fig. 1 can be expanded in terms
of Goldstone diagrams. In fact, this is  an infinite
series expansion, which contains RPA-diagrams, self-energy
ones, etc. In
Fig. 2, we show some of these terms, where
the pion decays into a particle-hole pair
($ph$), a $\Delta$-hole pair ($\Delta h$), etc. Eventually,
the $ph$ or $\Delta h$ can propagate within the nuclear medium.
Each term has its own poles, and it is
possible to evaluate the mesonic
and nonmesonic contributions to the $\Lambda$ decay separately.
For example, when the pion self energy is neglected in Eq.~(\ref{pprop}),
a pole in the meson propagator
(which means that the meson is on the mass shell)
gives a  contribution to the mesonic
rate. On the other hand, in the nonmesonic decay rate
the mesons are off the mass shell.

 So far we have pointed out  that the PPM is
the sum of an infinite series of diagrams, containing both mesonic
and nonmesonic contributions.
Specific formulas and further details, can be found in
Ref. \cite{os85}. From now on, we concentrate on the nonmesonic
decay rate $\Gamma_{NM}$, considering
only the contribution of the second diagrams displayed in Fig. 2, which is
evaluated by means of the Goldstone rules.
To obtain an analytical  expression for $\Gamma_{NM}$
we should first specify the one pion
exchange transition potential  in momentum space
\begin{equation}
\label{ppot}
V_{\pi}(q) =  G_F m_{\pi}^2 \; \frac{g_{NN \pi}}{2 M} \, F^2_{\pi}(q) \,
\left( \hat{A} + \frac{\hat{B}}{2 \bar{M}} \; \v{\sigma_1 \, \cdot \,
q} \right) \; \frac{\v{\sigma_2 \, \cdot \, q}}{q_0^2 - \v{q}^2 -
m_{\pi}^2},
\end{equation}
which is  the nonrelativistic reduction of
Eqs.~(\ref{Hamilt}) and (\ref{Hamilts}). Here $\v{q}$ is the momentum
carried by the pion and $M$
is the nucleon mass. Note that we have added the form
factor $F_{\pi}(q)$. The operators $\hat{A}$ and
$\hat{B}$, which contains the isospin dependence of the
potential, are
\begin{eqnarray}
\label{ipion}
\hat{A} = A_{\pi} \v{\tau_1 \cdot \tau_2}, \\
\hat{B} = B_{\pi} \v{\tau_1 \cdot \tau_2}.
\end{eqnarray}
Let us consider analytical expression for nonmesonic decay rate
from the above mentioned diagram. It is convenient to distinguish
between the proton ($t_h=1/2$) and neutron ($t_h=-1/2$) decay rates
\begin{eqnarray}
\label{gamdirpi0}
\Gamma_{t_h}^{dir}(\v{k},k_F)  & =
& -2 \, Im \, \int \frac{d^4 \, q}{(2 \pi)^4} \,
\int \frac{d^4 \, \kappa}{(2 \pi)^4} \;  G_N(\kappa+q/2) \; G_N(\kappa-q/2)
\nonumber \\
&& G_N(k-q) \; \frac{1}{4}
\sum_{s_pt_ps_{p'}t_{p'}s_\Lambda s_h} \;
|\bra{s_pt_ps_{p'}t_{p'}} V_{\pi}(q)\ket{s_\Lambda t_\Lambda s_ht_h}|^{2},
\end{eqnarray}
where the $s$'s and the $t$'s stand for the spin and isospin quantum numbers (see Fig. 2).
When performing the energy integrations one
keeps only the two particles-one hole ($2p1h$) cut, which gives the nonmesonic
character to the transition rate.
For the neutron decay we obtain,
\begin{eqnarray}
\label{gamdirpi}
\Gamma_{n}^{dir}(\v{k},k_F)  & =
&  \left(G_F m_{\pi}^2 \; \frac{g_{NN \pi}}{2 M} \;
\frac{1}{2 \pi}\right )^2 \; \int d \v{q} \;
\theta(q_0) \theta(|\v{k-q}|-k_F) \nonumber \\
& &  F_{\pi}^2(\v{q})\left [ A_{\pi}^2 +
\v{q}^2 \left(\frac{B_{\pi}}{2 \bar{M}}\right)^2 \right]
\; \frac{\v{q}^2}{(q_0^2 - \v{q}^2 -
m_{\pi}^2)^2} \; {\cal U}(q_0,\v{q}),
\end{eqnarray}
while $\Gamma_{p}^{dir}(\v{k},k_F)=5\Gamma_{n}^{dir}(\v{k},k_F)$.
Here $q_0=k_0 - E_N(\v{k-q}) - V_N$, with $k_0$ being the energy
of the $\Lambda$ and
\begin{eqnarray}
\label{lind}
{\cal U}(q_0,\v{q}) & = & \frac{2}{(2 \pi)^2} \; \int d \v{\kappa}
\; \; \theta(|\v{\kappa}+\frac{\v{q}}{2}|-k_F) \theta(k_F-
|\v{\kappa}-\frac{\v{q}}{2}|)  \nonumber \\
& & \delta(q_0 - (E_N(\v{\kappa}+\frac{\v{q}}{2}) - E_N( \v{\kappa}
- \frac{ \v{q} }{2})))
\end{eqnarray}
is the Lindhard function.
The corresponding $\Gamma^{dir}_{n,p}$ are obtained  from  Eqs. (\ref{decwpar2})
and (\ref{decwpar3}).

At this point, we would like to discuss the limitations
of the PPM.
The first one refers to the incorporation
of other mesons beyond the pion. By construction, in
the PPM the only weak vertex is the $\Lambda N \pi$
one. The incorporation of other mesons is certainly
important for the nonmesonic decay width, but it is not
compatible with Eq.~(\ref{self1}). Note that the naive attempt to
solve this problem by replacing the pion propagator
$G_{\pi}$ with the sum of other mesons propagators
($G_{\pi} \rightarrow G_{\pi}\, + \, G_{\rho}\, + \,
G_{\eta} \, + \, ...$), fails to incorporate
interference terms between mesons. The second point
refers to the exchange terms, which
are obviously  not
included in the PPM, as all diagrams are originated from
the expansion of the dressed pion propagator of Eq. (\ref{pprop}).
In this  work we develop a nuclear matter
scheme, which overcome both
just mentioned difficulties. We will limit our attention to the diagrams displayed
in the	 Fig. 3, where the second graph of Fig. 2 is redrawn
in the part $a)$,
while in the part $b)$ we show it the corresponding
exchange contribution.

We use the standard
strangeness-changing weak $\Lambda N\to NN$ transition potential which
involves  the exchange of the complete pseudoscalar and vector meson octets
($\pi,\eta,K,\rho,\omega,K^*$).
It was taken from Ref.~\cite{pa95} and the
explicit expressions  are listed in
Appendix A. The incorporation of the short range correlations (SRC) is
explained in Appendix B.
For the sake of convenience,  the total
transition potential
is written as

\begin{equation}
\label{2.12}
V_{SRC} (q) = \sum_{\tau=0,1} {\cal O}_\tau
{\cal V}_\tau(q),~~~~~ {\cal O}_\tau=\left\{
\begin{array}{c}1\\
  \v{\tau}_1 \cdot \v{\tau}_2
\end{array}\right.,
\end{equation}

where
\begin{eqnarray}
\label{allmespot}
{\cal V}_{\tau}(q) &
= &  (G_F m_{\pi}^2)  \; \{
S_{\tau}(q)  \; \v{\sigma_1 \cdot \hat{q}} +
S'_{\tau}(q)  \; \v{\sigma_2 \cdot \hat{q}} +
P_{L, \tau}(q)	\v{\sigma_1 \cdot \hat{q}} \; \v{\sigma_2 \cdot
\hat{q}} + P_{C, \tau}(q)  +  \nonumber \\
& & +  P_{T, \tau}(q)  (\v{\sigma_1 \times \hat{q}})
\cdot  (\v{\sigma_2 \times \hat{q}}) + i S_{V, \tau}(q)
\v{(\sigma_1 \times \sigma_2) \cdot \hat{q}} \}.
\end{eqnarray}
The  quantities  $S_{\tau}(q)$, $S'_{\tau}(q)$, $P_{L, \tau}(q)$,
$P_{C,\tau}(q)$, $P_{T, \tau}(q)$ and $S_{V, \tau}(q)$
contain
the SRC and  are also given in Appendix B.
The values $\tau=0,1$ stand for the isoscalar
and isovector parts of the interaction, respectively.

The corresponding transition rate is the sum of diagrams $a)$
and $b)$ in Fig. 3, $\Gamma_{n,p}=\Gamma^{dir}_{n,p}+\Gamma^{exch}_{n,p}$.
Each one is obtained by modifying Eq.~(\ref{gamdirpi0}) as follows.
The direct contribution, $\Gamma^{dir}_{n,p}(\v{k},k_F)$,
is obtained from  the replacement
$V_{\pi}(q)  \, \rightarrow \, V_{SRC}(q)$, while for the exchange contribution,
$\Gamma^{exch}_{n,p}(\v{k},k_F)$,
one substitutes
$|\bra{s_pt_ps_{p'}t_{p'}} V_{\pi}(q)\ket{s_\Lambda t_\Lambda s_ht_h}|^{2}$
by
$$(-)\bra{s_pt_ps_{p'}t_{p'}} V_{SRC}(q)\ket{s_\Lambda t_\Lambda s_ht_h}^{*}
\bra{s_{p'}t_{p'}s_pt_p} V_{SRC}(Q)\ket{s_\Lambda t_\Lambda s_ht_h},$$
 where the minus sign
comes from  the crossing  of the fermionic lines
and $Q \equiv k - \kappa - q/2$.

In order to perform the summation on spin and isospin quantum numbers it is
convenient to rewrite the transition rates in the form
\begin{equation}
\label{gammtot}
\Gamma_{n,p}^{dir,exch}(\v{k},k_F) = \sum_{\tau=0,1}
{\cal T}^{dir,exch}_{n,p; \; \tau \tau'} \; \;
\widetilde{\Gamma}_{\tau \, \tau'}^{dir, exch}(\v{k},k_F)  \;
\end{equation}
where
\begin{eqnarray}
\label{sumisosp}
{\cal T}^{dir}_{t_h; \, \tau \tau'} & = & \sum_{t_p, t_{p'}}
\bra{t_\Lambda t_h} {\cal O}_\tau \ket{t_pt_{p'}}
\bra{t_pt_{p'}} {\cal O}_{\tau'} \ket{t_\Lambda t_h} \nonumber \\
{\cal T}^{exch}_{t_h; \, \tau \tau'} & = & \sum_{t_p, t_{p'}}
\bra{t_\Lambda t_h} {\cal O}_\tau \ket{t_pt_{p'}}
\bra{t_{p'}t_p} {\cal O}_{\tau'} \ket{t_\Lambda t_h}
\end{eqnarray}
The partial
decay widths are defined as,
\begin{eqnarray}
\label{gamdir}
\widetilde{\Gamma}_{\tau \, \tau'}^{dir}(\v{k},k_F)  & =
& (G_F m_{\pi}^2)^2 \frac{1}{(2 \pi)^5} \int  \int d
\v{q} d \v{\kappa} \; {\cal S}^{dir}_{\tau \tau'}(q)  \;
\theta(q_0) \theta(|\v{k-q}|-k_F) \nonumber \\
& & \theta(|\v{\kappa}+\frac{\v{q}}{2}|-k_F) \theta(k_F-
|\v{\kappa}-\frac{\v{q}}{2}|)  \;
\delta(q_0 - (E_N(\v{\kappa}+\frac{\v{q}}{2}) - E_N( \v{\kappa}
- \frac{ \v{q} }{2})))
\end{eqnarray}
for the direct contribution, and
\begin{eqnarray}
\label{gamexc}
\widetilde{\Gamma}_{\tau \, \tau'}^{exch}(\v{k},k_F) & =
& (G_F m_{\pi}^2)^2 \frac{1}{(2 \pi)^5} \int  \int d
\v{q} d \v{\kappa} \; {\cal S}^{exch}_{\tau \tau'}(q,Q)  \;
\theta(q_0) \theta(|\v{k-q}|-k_F) \nonumber \\
& & \theta(|\v{\kappa}+\frac{\v{q}}{2}|-k_F) \theta(k_F-
|\v{\kappa}-\frac{\v{q}}{2}|)  \;
\delta(q_0 - (E_N(\v{\kappa}+\frac{\v{q}}{2}) - E_N( \v{\kappa}
- \frac{ \v{q} }{2})))
\end{eqnarray}
for the exchange one, where $\v{Q}=\v{k}-\v{\kappa}-\v{q}/2$ and
$Q_0=k_0-E_N(\v{\kappa} +
\v{q}/2) - V_N$. The integration over $\v{\kappa}$ in
Eq.~(\ref{gamdir}) is factorized as the Lindhard function, which
simplifies the evaluation of $\widetilde{\Gamma}^{dir}_{\tau \, \tau'}$.
Summation over spin is already performed in ${\cal
S}^{dir}_{\tau,\tau'}(q)$ and ${\cal S}^{exch}_{\tau,\tau'}(q,Q)$,
which are defined as,
\begin{eqnarray}
\label{tdir}
{\cal S}^{dir}_{\tau \tau'}(q) & = &  4 \; \{
S_{\tau}(q) S_{\tau'}(q) + S'_{\tau}(q) S'_{\tau'}(q)
 + P_{L, \tau}(q)  P_{L, \tau'}(q)  +  P_{C, \tau}(q) P_{C,
\tau'}(q) + \nonumber \\
 & & + 2  \, P_{T, \tau}(q)  P_{T, \tau'}(q) + 2 \, S_{V,
\tau}(q) S_{V, \tau'}(q) \}
\end{eqnarray}
and
\begin{eqnarray}
\label{texc}
{\cal S}^{exch}_{\tau \tau'}(q,Q) & = &
(\hat{\v{q}} \cdot \hat{\v{Q}})  \textsf{S}_{\tau}(q) \textsf{S}_{\tau'}(Q)
 + ( 2 (\hat{\v{q}} \cdot \hat{\v{Q}})^2 - 1)  P_{L,
\tau}(q)  P_{L, \tau'}(Q)  + \nonumber \\
 & & + P_{C, \tau}(q) P_{C, \tau'}(Q)
 + 2 ((\hat{\v{q}} \cdot \hat{\v{Q}})^2 - 1)   P_{T,
\tau}(q)  P_{T, \tau'}(Q)- \nonumber \\
& & - 2 (\hat{\v{q}} \cdot \hat{\v{Q}})^2 (P_{L,
\tau}(q) P_{T, \tau'}(Q) + P_{L, \tau}(Q) P_{T,
\tau'}(q)).
\end{eqnarray}
where
\begin{eqnarray}
\label{sexc}
\textsf{S}_{\tau}(q) \textsf{S}_{\tau'}(Q) & = &
(S_{\tau}(q) + S'_{\tau}(q))(S_{\tau'}(Q)
+ S'_{\tau'}(Q))
\nonumber \\
& & + 2 (S_{\tau}(q) S_{V, \, \tau'}(Q) +
S_{V, \, \tau}(q) S_{\tau'}(Q)) \nonumber \\
& & - 2 (S'_{\tau}(q) S_{V, \, \tau'}(Q) +
S_{V, \, \tau}(q) S'_{\tau'}(Q)).
\end{eqnarray}
The partial widths $\widetilde{\Gamma}^{dir, \,
exch}_{\tau \, \tau'}(\v{k},k_F) $ depend on the
momentum of $\Lambda$ and on the Fermi momentum, $k_F$.
The $k_F$-dependence is eliminated by means of the
LDA, as shown in Eq.~(\ref{decwpar2}), {\em i.e.,}

\begin{equation}
\label{pdec}
\widetilde{\Gamma}^{dir, \, exch}_{\tau \, \tau'} \equiv \int d \v{k} \,
| \psi_{\Lambda}(\v{k})|^2 \, \widetilde{\Gamma}_{\tau \, \tau'}^{dir \,
exch}(\v{k}).
\end{equation}
The final result from Eq. (\ref{gammtot}) is
respectively,
\begin{eqnarray}
\label{decnp}
\Gamma_n & = &	\widetilde{\Gamma}^{dir}_{11} -
\widetilde{\Gamma}^{exch}_{11} +  \widetilde{\Gamma}^{dir}_{00} -
\widetilde{\Gamma}^{exch}_{00} + \widetilde{\Gamma}^{dir}_{01}-
\widetilde{\Gamma}^{exch}_{01}+\widetilde{\Gamma}^{dir}_{10} -
\widetilde{\Gamma}^{exch}_{10}
\nonumber \\
\Gamma_p & = & 5 \, \widetilde{\Gamma}^{dir}_{11} + 4 \,
\widetilde{\Gamma}^{exch}_{11} +  \widetilde{\Gamma}^{dir}_{00} -
(\widetilde{\Gamma}^{dir}_{01}+\widetilde{\Gamma}^{dir}_{10}+2 \,
\widetilde{\Gamma}^{exch}_{01}+ 2 \, \widetilde{\Gamma}^{exch}_{10}).
\end{eqnarray}
Note that $\Gamma_{n/p}=1/5$, when
only the direct isovector contributions are considered.
In the next section we give numerical results and also
analyze the importance of different terms entering into
our scheme.

In the evaluation of exchange term (diagram $b$ of Fig. 3)  Jido,
Oset and Palomar \cite{ji01} have approximated the momentum
$\v{Q}$  by $-\v{q}$, which greatly simplifies the calculation.
This implies that the exchange
terms in Ref. \cite{ji01} are approximated by direct ones, but
with the spin-isospin factors corresponding to actual exchange
diagrams. Simultaneously, they consider that
the $\Lambda$ carries a non-vanishing $\v{k}$-momentum in
 both the direct and the exchange terms.
This last point is somehow contradictory with
$\v{Q} \approx -\v{q}$, as
the later  approximation is based on: ${\it i}$)
the hyperon is considered to be at rest ($\v{k}=0$), and ${\it
ii}$) the momentum of the hole is
neglected ($\v{\kappa}-\v{q}/2=0$).
 To arrive to the same simplification from our
scheme, one simply replaces $\v{Q}$ by
$-\v{q}$ and $Q_0$ by $q_0$ in Eq.~(\ref{texc}), which makes the quantity ${\cal
S}^{exch}_{\tau,\tau'}$ to depend only on $q$. As a further
consequence, all factors $(\hat{\v{q}} \cdot \hat{\v{Q}})$ goes to
-1, the term $P_{T, \tau}(q)  P_{T,\tau'}(Q)$ disappears and
$\v{\kappa}$-integral in Eq.~(\ref{gamexc}) reduces to the
Lindhard function.

\newpage
\section{RESULTS}
\label{RESULT}

In this section we give numerical values for the nonmesonic
$\Lambda$-decay width. All calculations were done in nuclear
matter with  the transition potential presented in the last
section and in Appendices A and B. The results for
$^{12}_{\Lambda}C$ comes from the LDA.
The multiple integrations have been performed using a Monte Carlo
technique. As mentioned in Sect. II,
the hyperon is assumed to be in the $1s_{1/2}$ orbit of  a
harmonic oscillator well with frequency $\hbar \omega = (45
A^{-1/3} \, - \, 25 A^{-2/3})$ MeV.

In order to analyze the importance of the exchange terms we show
in Table I the numerical results for the neutron and proton decay
widths. Two results are displayed for the total (direct plus
exchange) decay rates,\\
{\it Calculations I}: The simplification explained at the end of the last
section has been implemented for the exchange term.
\\
{\it Calculations II}: The exchange contribution is evaluated in the exact way.\\
We start with the results for the OPE and then we add one by one the
contributions of the remainder
 transition potentials.  As expected the exchange terms are quite important.
  Furthermore, one sees that the total transition rates strongly depend on the
way these terms are  evaluated.
 The final result shows that the exchange terms increase
the value of $\Gamma_n$ while it has the opposite effect over
$\Gamma_p$, improving the ratio $\Gamma_{n/p}$.

We have paid some attention to the role of the
$\rho$-meson and the relative contributions of the parity
violating ($PV$) and parity conserving ($PC$) decay widths.
Since the work of McKeller and
Gibson \cite{mc84}, there was a controversy referring to the
importance of  this meson. Yet in  the work of
Parre\~no {\it et al.} \cite{pa95}, it was established that
when the $\rho$-meson is added to the pion  the
total rate is reduced by about 10-15 \%.
 In Table II, we compare our nuclear matter results
with the finite nucleus calculation of Barbero {\it et al.} \cite{ba02}.
In the present calculation, the reduction of total
rate is slightly bigger than in \cite{pa95,ba02},
although  the overall agreement is rather good.
It is worth noting that the present values for $\Gamma^{PV}_{n,p}$ and
$\Gamma^{PC}_{n,p}$ differ from those of finite nucleus in the case of	the
pion, but for the $\rho$-meson the  agreement is satisfactory.
While in  finite nucleus
calculations   the pion PV contribution is about 40\%
of the total $\pi$-meson decay width \cite{pa97,ba02},
we get that it is only of about
23\%. Note however, that the latter percentage is appreciable
larger that in previous
 nuclear matter
estimates:  it is  negligible  in  Ref.~\cite{mc84} and
of the order of $ 15$\% in  Ref. \cite{du86}.

In Table III, we analyze the role of the $K$-meson, which
improves the value of $\Gamma_n/\Gamma_p$. We compare
our results with  the nuclear matter
calculations done in Refs.~\cite{sa00,ji01}. In the first work
a partial wave
expansion of the nuclear matter plane waves is done. The decay width is
also evaluated in an approximate way: the summation over
momentum of the two outgoing particles is performed with no
restrictions (which means that they could take values below the
Fermi momentum). The values for $\Gamma_n/\Gamma_p$ from Ref.~\cite{sa00}
are in agreement with ours, but the individual transitions rates,
$\Gamma_{n, \, p}$, are bigger.
Regarding the second work, it should be stressed that
the differences with  our {\it Calculation I} are:
1) in  Ref.~\cite{ji01}
are also included the RPA correlations, and 2) the effective interaction is
somewhat different. The second effect turns out to be  the most relevant, as
can be seen from the {\it Calculation I'}, where
 $\Gamma_{n}$ and $\Gamma_{p}$	are evaluated by employing both
the approximation and  the interaction from Ref.~\cite{ji01}, but
without the RPA correlations.
Finally, note that   the inclusion of the kaon	increases the ratio
$\Gamma_n/\Gamma_p$ within the	{\it
Calculation II} as well.

In Table IV we
compare our results
for the full OME with those of Ref.~\cite{ba02}. They
 are quite similar, except
 for the vector mesons $\omega$ and $K^*$.
One should keep in mind that the finite nucleus formalism
of Ref.~\cite{ba02} has notable
differences compared with the present nuclear matter model. Among the
sources of difference, we can mention that we employ
plane waves for both the incoming and outgoing wave
functions, while in Ref.~\cite{ba02} harmonic oscillator wave
functions for the incoming particles are used and the
outgoing nucleons are expanded into partial waves.

Before ending this section we must call attention on the RPA-correlations. Many
nuclear matter calculations dress the mesons
propagators with RPA-type correlations. More precisely, very frequently only the
direct RPA terms  are considered, an estimation which is usually
called ring approximation (RA).  Within this framework a strong dependence of the
total nonmesonic decay width on the  Landau-Migdal coupling $g'$
 has been reported recently  in Ref.~\cite{al00b}
(see Fig. 3 of this work).
 Yet, it is well  known that the RA  leads
to significantly different results for the
electron scattering strength function than the full RPA
\cite{ba96,ba99}.
Thus,  we
consider that it is encouraging to explore  the consequences of the full RPA
on  the $\Lambda$-decay, which
certainly is a complex issue and is beyond scope of the present work.

 It is worthwhile to say a few	words on the final state
interactions (FSI), which
is a very general denomination for
all kind of interactions between the two outgoing particles.
Parre\~no and Ramos ~\cite{pa02} have treated them recently through the solution of a
$T$-matrix using realistic NN interactions. Their results show that the
FSI demand this kind of calculations over the phenomenological approach,
which has been used in the present work.

As a final comment for this section, we wish to restate that there are
several methods  to relate the nuclear matter results
to experimental data.
 Besides
the LDA, we can mention the use of an effective Fermi
momentum \cite{am94}, and the employment of a diffused Fermi
surface \cite{ba01}. These last two approximations have been successfully
employed  in the context of the electron-nucleus
quasi-elastic scattering.
However, they lead to non-physical results for the mesonic decay,
which is totally forbidden when the first method is employed
and becomes artificially big when the second one is used \cite{os98}.
These elements suggest that the LDA is a
more adequate approximation for hypernuclei decays.

\newpage

\section{CONCLUSIONS}
\label{CONCLU}
A  nuclear matter scheme for calculating the
nonmesonic $\Lambda$-decay width has been presented,
with explicit inclusion of exchange terms. It has been assumed that
the transition is triggered by the full pseudoscalar-vector meson octet,
with the corresponding form
factors and short range correlations. To evaluate  the	decay rate
of $^{12}_{\Lambda}C$ the LDA has been employed as well.
Our numerical results were compared with finite nucleus ones and,
except for the $\omega$ and $K^{*}$ mesons,
 good agreement was obtained.

At variance with finite nucleus
calculations,  the exchange terms are not always taken into account
in the	nuclear matter studies. In fact, the last ones can be classified in two
groups, depending on whether
the  partial wave expansion of the
nuclear matter plane waves is performed or not. In the first case,
the Pauli principle is considered, but the $\Lambda$
is taken to  be at rest and it is implicitly assumed that the exchange
term  carries the same
momentum as the direct one (for details see Ref. \cite{mc84}).
Within the second group  the most relevant formalism is, in our opinion,
 the PPM put forward by Oset and Salcedo \cite{os85}.
The majority of works done within this model do not
include the exchange term, which implies  a separate and more complex calculation.
 An exception is Ref.~\cite{ji01}, where they  are
incorporated in an approximate way (as stated at the end of Section II).
Contrarily, we have evaluated them exactly, arriving to the conclusion
that they are not only important but that they should also be calculated accurately.

Our numerical results agree fairly well with those obtained
within the shell model	 framework \cite{pa97,ba02}. Same as
in these works, we are able to reproduce the data
for the total nonmesonic decay width: $\Gamma_{NM}^{exp}\sim\Gamma_0$ \cite{Mo74},
but not that for the $n/p$ ratio
: $\Gamma_{n/p}^{exp}= 1.17^{+0.09+0.22}_{-0.08-0.18}$ \cite{Ha02}.
This suggests that some others relevant physical ingredients are still missing.
In this sense, our nuclear
matter formalism  is particularly suitable
for: 1)   analyzing the RPA correlations,  and 2) the inclusion
of the $\Lambda NN \rightarrow NNN$ decay \cite{al91,ra94,ra97,al00b}.

\newpage

\acknowledgements

The authors are fellows of the CONICET	Argentina, and acknowledge  the support of
ANPCyT (Argentina) under grant BID 1201/OC-AR (PICT 03-04296).
 One of us (E.B.) would like to thank
A. Ramos for very helpful and illuminating discussions, and for a careful and
critical reading of the manuscript.

\newpage

\section*{APPENDIX A}
\label{APPENDA}
In this appendix we show explicit expressions for the
($\eta$, $K$, $\rho$, $\omega$ and $K^*$)-$N \Lambda
\rightarrow NN$ transition potential. The formulation
was taken from \cite{pa95}, while the values of the
different coupling constants and cutoff parameters
appearing in the transition
potential were taken from \cite{na77}. Weak couplings
are in units of $G_F \, m_{\pi}^2$.

For the pseudoscalar mesons we have expressions similar
to Eq.~(\ref{ppot}) but making the following replacements,
\begin{eqnarray}
\label{eta}
g_{NN \pi} & \rightarrow & g_{NN \eta}, \nonumber \\
m_{\pi} & \rightarrow & m_{\eta}, \nonumber \\
\hat{A} & \rightarrow & A_{\eta}, \nonumber \\
\hat{B} & \rightarrow & B_{\eta},
\end{eqnarray}
for the exchange of the isoscalar $\eta$-meson and
\begin{eqnarray}
\label{kaon}
g_{NN \pi} & \rightarrow & g_{\Lambda N K}, \nonumber \\
m_{\pi} & \rightarrow & m_K, \nonumber \\
\hat{A} & \rightarrow & (\frac{C_K^{PV}}{2} + D_K^{PV}
+ \frac{C_K^{PV}}{2} \; \v{\tau_1 \cdot \tau_2})
\frac{M}{\bar{M}}, \nonumber \\
\hat{B} & \rightarrow & -(\frac{C_K^{PC}}{2} + D_K^{PC}
+ \frac{C_K^{PC}}{2} \; \v{\tau_1 \cdot \tau_2})
\end{eqnarray}
together with the exchange of index 1 and 2 in spin,
for the isodoublet kaon. We employ, $g_{NN \eta}=6.4$,
$A_{\eta}=1.8$, $B_{\eta}=-14.3$ and $\Lambda_{\eta}=1.3$
GeV, for the $\eta$-meson. For the $K$ meson, $g_{\Lambda N K}
=-14.1$, $C_K^{PV}=0.76$, $C_K^{PC}=-18.9$, $D_K^{PV}=2.09$,
$D_K^{PC}=6.63$ and $\Lambda_{K}=1.2$ GeV.
In the case of vector mesons, we start with the $\rho$-meson,
\begin{eqnarray}
\label{rho}
V_{\rho}(q) & = & G_F m_{\pi}^2 \; ( \; F_1 \hat{\alpha} -
\frac{(\hat{\alpha}+\hat{\beta})(F_1+F_2)}{4 M \bar{M}}
(\v{\sigma_1} \times \v{q}) \cdot (\v{\sigma_2} \times \v{q}) -
\nonumber \\
& & - i \hat{\varepsilon} \frac{F_1+F_2}{2 M}
(\v{\sigma_1} \times \v{\sigma_2}) \cdot \v{q} \; ) \;
\frac{1}{q_0^2 - \v{q}^2 - m_{\rho}^2}
\end{eqnarray}
where $F_1=g_{NN \rho}^V$ and $F_1=g_{NN \rho}^T$ and the
operators $\hat{\alpha}$, $\hat{\beta}$ and $\hat{\varepsilon}$ are,
\begin{eqnarray}
\label{irho}
\hat{\alpha} & = & \alpha_{\rho} \, \v{\tau_1 \cdot \tau_2},
\nonumber \\
\hat{\beta} & = & \beta_{\rho} \, \v{\tau_1 \cdot \tau_2},
\nonumber \\
\hat{\varepsilon} & = & \varepsilon_{\rho} \, \v{\tau_1 \cdot \tau_2},
\end{eqnarray}
with $g_{NN \rho}^V=3.16$, $g_{NN \rho}^T=13.3$, $\alpha_{\rho}=-3.50$,
$\beta_{\rho}=-6.11$, $\varepsilon_{\rho}=1.09$
and $\Lambda_{\rho}=1.4$ GeV.
Finally, to obtain the $\omega$ and $K^*$ terms, one has to make the
following substitutions in Eq.~(\ref{rho}),
\begin{eqnarray}
\label{omega}
m_{\rho} & \rightarrow & m_{\omega}, \nonumber \\
F_1 & \rightarrow & g_{NN \omega}^V, \nonumber \\
F_2 & \rightarrow & g_{NN \omega}^T,\nonumber \\
\hat{\alpha} & \rightarrow & \alpha_{\omega}, \nonumber \\
\hat{\beta} & \rightarrow & \beta_{\omega}, \nonumber \\
\hat{\varepsilon} & \rightarrow & \varepsilon_{\omega}
\end{eqnarray}
and
\begin{eqnarray}
\label{kaone}
m_{\rho} & \rightarrow & m_{K^*}, \nonumber \\
F_1 & \rightarrow & g_{\Lambda N K^*}^V, \nonumber \\
F_2 & \rightarrow & g_{\Lambda N K^*}^T, \nonumber \\
\hat{\alpha}  & \rightarrow & \frac{C_{K^*}^{PC, \, V}}{2}
+ D_{K^*}^{PC, \, V}
+ \frac{C_{K^*}^{PV, \, V}}{2} \; \v{\tau_1 \cdot \tau_2}
\nonumber \\
\hat{\beta}  & \rightarrow & \frac{C_{K^*}^{PC, \, T}}{2}
+ D_{K^*}^{PC, \, T}
+ \frac{C_{K^*}^{PC, \, T}}{2} \; \v{\tau_1 \cdot \tau_2}
\nonumber \\
\hat{\varepsilon}  & \rightarrow & (\frac{C_{K^*}^{PV}}{2}
+ D_{K^*}^{PV}
+ \frac{C_{K^*}^{PV}}{2} \; \v{\tau_1 \cdot \tau_2})
\frac{M}{\bar{M}},
\end{eqnarray}
with $g_{NN \omega}^V=10.5$, $g_{NN \omega}^T=3.22$,
$\alpha_{\omega}=-3.69$, $\beta_{\omega}=-8.04$,
$\varepsilon_{\omega}=-1.33$, $\Lambda_{\omega}=1.50$ GeV,
$g_{\Lambda N K^*}^V=-5.47$, $g_{\Lambda N K^*}^T=-11.9$
$C_{K^*}^{PC, \, V}=-3.61$, $C_{K^*}^{PC, \, T}=-17.9$,
$C_{K^*}^{PV}=-4.48$, $D_{K^*}^{PC, \, V}=-4.89$,
$D_{K^*}^{PC, \, T}=9.30$, $D_{K^*}^{PV}=0.60$ and
$\Lambda_{K^*}=2.20$ GeV.

\newpage

\section*{APPENDIX B}
\label{APPENDB}

In momentum space the short range correlated (SRC)
transition potential is obtained as,
\begin{equation}
\label{srcp}
V_{SRC}(\v{q}) \, = \, V(\v{q}) \, - \, \int \frac{d \v{p}}{(2 \pi)^3}
\tilde{\xi}(|\v{p} + \v{q}|) \, V(\v{p})
\end{equation}
where,
\begin{equation}
\label{cfunp}
\tilde{\xi}(p) = \frac{2 \pi^2}{q_c^2} \, \delta(p-q_c)
\end{equation}
is the correlation function in momentum space. We have used $q_c = 780$.
As an example let us show the result
of Eq.~(\ref{srcp}) with the central part of the parity conserving
one pion exchange potential, which we write in a simplify manner as,
\begin{equation}
\label{pionc}
V_{\pi}^C(\v{q})  =  C_{\pi} \; \frac{\v{q}^2}{\v{q}^2+m_{\pi}^2}
\v{\sigma_1} \cdot \v{\sigma_2} \;
\v{\tau_1} \cdot \v{\tau_2}
\end{equation}
with $C_{\pi} = - G_F m_{\pi}^2 \; \frac{g_{NN \pi}}{2 M} \;
\frac{B_{\pi}}{2 \bar{M}}$. Using this potential in
Eq.~(\ref{srcp}) we obtain,
\begin{equation}
\label{srcpc1}
V^{SRC, \, C}(\v{q}) \, = \, V^C(\v{q}) \, - \,
C_{\pi} \frac{1}{2} \{ 2 + \frac{m_{\pi}^2}{2 q_c |\v{q}|}
\ln |\frac{q_c^2 + m_{\pi}^2 + \v{q}^2 - 2 q_c |\v{q}|}
      {q_c^2 + m_{\pi}^2 + \v{q}^2 + 2 q_c |\v{q}|}| \}
\v{\sigma_1} \cdot \v{\sigma_2} \;
\v{\tau_1} \cdot \v{\tau_2}
\end{equation}
if we call $\kappa = 2 q_c |\v{q}|/(q_c^2 + m_{\pi}^2 + \v{q}^2)$ and
now we use,
\begin{equation}
\ln (1 + \kappa) \approx \kappa
\end{equation}
we finally obtain,
\begin{equation}
\label{srcpc2}
V^{SRC, \, C}(\v{q}) \, = \, V^C(\v{q}) \, - \,
C_{\pi} \frac{q_c^2 + \v{q}^2}{q_c^2 + m_{\pi}^2 + \v{q}^2}
\v{\sigma_1} \cdot \v{\sigma_2} \;
\v{\tau_1} \cdot \v{\tau_2}
\end{equation}
which means that the contribution steaming from the second term of
the r.h.s of Eq.~(\ref{srcp}) is simply $V^C(\v{q}^2 \rightarrow
q_c^2 + \v{q}^2)$. The procedure is analogous for rest of the interaction.
We present now the final results of the short range correlated
($\pi$ + $\eta$ + $K$ + $\rho$ + $\omega$ + $K^*$)-transition potential.
First, we define the following quantities,
\begin{eqnarray}
\label{prop}
{\cal W}_{\pi}(q) & = & \frac{g_{NN \pi}}{2 M} \; \frac{B_{\pi}}{2
\bar{M}} F_{\pi}^2(q) G_{\pi}(q) \nonumber \\
{\cal W}_{\pi}^S(q) & = & \frac{g_{NN \pi}}{2 M} \; A_{\pi} \,
F_{\pi}^2(q) G_{\pi}(q) \nonumber \\
{\cal W}_{\eta}(q) & = & \frac{g_{NN \eta}}{2 M} \;
\frac{B_{\eta}}{2 \bar{M}} F_{\eta}^2(q) G_{\eta}(q) \nonumber
\\
{\cal W}_{\eta}^S(q) & = & \frac{g_{NN \eta}}{2 M} \; A_{\eta} \,
F_{\eta}^2(q) G_{\eta}(q) \nonumber \\
{\cal W}_K^0(q) & = & - \frac{g_{\Lambda N K}}{2 M} \; \frac{1}{2
\bar{M}} (\frac{C_K^{PC}}{2} + D_K^{PC}) \frac{M}{\bar{M}}
F_K^2(q) G_{K}(q) \nonumber \\
{\cal W}_K^{S, \, 0}(q) & = & \frac{g_{\Lambda N K}}{2 M} \;
(\frac{C_K^{PV}}{2}  + D_K^{PV}) \, F_K^2(q) G_{K}(q) \nonumber \\
{\cal W}_K^1(q) & = & - \frac{g_{\Lambda N K}}{2 M} \; \frac{1}{2
\bar{M}} \frac{C_K^{PC}}{2} \frac{M}{\bar{M}} F_K^2(q) G_{K}(q)
\nonumber \\
{\cal W}_K^{S, \, 1}(q) & = & \frac{g_{\Lambda N K}}{2 M} \;
\frac{C_K^{PV}}{2} \, F_K^2(q) G_{K}(q) \nonumber \\
{\cal W}_{\rho}^C(q) & = & \alpha_{\rho} g_{NN \rho}^V
F_{\rho}^2(q) G_{\rho}(q) \nonumber \\
{\cal W}_{\rho}^T(q) & = & -\frac{(\alpha_{\rho}+\beta_{\rho})
(g_{NN \rho}^V+g_{NN \rho}^T)}{4 M \bar{M}} F_{\rho}^2(q)
G_{\rho}(q) \nonumber \\
{\cal W}_{\rho}^{PV}(q) & = & \frac{\varepsilon_{\rho} (g_{NN
\rho}^V+g_{NN \rho}^T)}{2 M} F_{\rho}^2(q) G_{\rho}(q) \nonumber
\\
{\cal W}_{\omega}^C(q) & = & \alpha_{\omega} g_{NN \omega}^V
F_{\omega}^2(q) G_{\omega}(q) \nonumber \\
{\cal W}_{\omega}^T(q) & = &
-\frac{(\alpha_{\omega}+\beta_{\omega}) (g_{NN \omega}^V+g_{NN
\omega}^T)}{4 M \bar{M}} F_{\omega}^2(q) G_{\omega}(q) \nonumber
\\
{\cal W}_{\omega}^{PV}(q) & = & \frac{\varepsilon_{\omega} (g_{NN
\omega}^V+g_{NN \omega}^T)}{2 M} F_{\omega}^2(q) G_{\omega}(q)
\nonumber \\
{\cal W}_{K^*}^{C, \, 0}(q) & = & (\frac{C_{K^*}^{PC, \, V}}{2} +
D_{K^*}^{PC, \, V}) g_{\Lambda N K^*}^V F_{K^*}^2(q) G_{K^*}(q)
\nonumber \\
{\cal W}_{K^*}^{T, \, 0}(q) & = & \frac{-1}{4 M \bar{M}}
(\frac{C_{K^*}^{PC, \, V}}{2} + D_{K^*}^{PC, \, V} +
\frac{C_{K^*}^{PC, \, T}}{2} + D_{K^*}^{PC, \, T}) (g_{\Lambda N
K^*}^V+g_{\Lambda N K^*}^T) F_{K^*}^2(q) G_{K^*}(q) \nonumber
\\
{\cal W}_{K^*}^{PV, \, 0}(q) & = & \frac{1}{2 M}
(\frac{C_{K^*}^{PV}}{2} + D_{K^*}^{PV}) (g_{\Lambda N
K^*}^V+g_{\Lambda N K^*}^T) F_{K^*}^2(q) G_{K^*}(q) \nonumber
\\
{\cal W}_{K^*}^{C, \, 1}(q) & = & \frac{C_{K^*}^{PC, \, V}}{2}
g_{\Lambda N K^*}^V F_{K^*}^2(q) G_{K^*}(q) \nonumber \\
{\cal W}_{K^*}^{T, \, 1}(q) & = & \frac{-1}{4 M \bar{M}}
(\frac{C_{K^*}^{PC, \, V}}{2} + \frac{C_{K^*}^{PC, \, T}}{2})
(g_{\Lambda N K^*}^V+g_{\Lambda N K^*}^T) F_{K^*}^2(q) G_{K^*}(q)
\nonumber \\
{\cal W}_{K^*}^{PV, \, 1}(q) & = & \frac{1}{2 M}
\frac{C_{K^*}^{PV}}{2} (g_{\Lambda N K^*}^V+g_{\Lambda N K^*}^T)
F_{K^*}^2(q) G_{K^*}(q) \nonumber \\
\end{eqnarray}
where,
\begin{equation}
G_i(q) = \frac{1}{q_0^2 - \v{q}^2 - m_{i}^2}
\end{equation}
we further introduce,
\begin{eqnarray}
\label{inti}
S_0 & = & ({\cal W}_{\eta}^S-\tilde{{\cal W}}_{\eta}^S) |\v{q}|
\nonumber \\
S_1 & = & ({\cal W}_{\pi}^S-\tilde{{\cal W}}_{\pi}^S) |\v{q}|
\nonumber \\
S'_0 & = & ({\cal W}_{K, \, 0}^S-\tilde{{\cal W}}_{K, \, 0}^S) |\v{q}|
\nonumber \\
S'_1 & = & ({\cal W}_{K, \, 1}^S-\tilde{{\cal W}}_{K, \, 1}^S) |\v{q}|
\nonumber \\
S_{V, 0} & = &
({\cal W}_{\omega}^{PV}-\tilde{{\cal W}}_{\omega}^{PV}
+{\cal W}_{K^*}^{PV, \, 0}-\tilde{{\cal W}}_{K^*}^{PV, \, 0}) |\v{q}|
\nonumber \\
S_{V, 1} & = &
({\cal W}_{\rho}^{PV}-\tilde{{\cal W}}_{\rho}^{PV}
+{\cal W}_{K^*}^{PV, \, 1}-\tilde{{\cal W}}_{K^*}^{PV, \, 1}) |\v{q}|
\nonumber \\
P_{L, 0} & = & q^2 ({\cal W}_{\eta} + {\cal W}_K^0)
- (q^2 + \frac{1}{3} q_c^2)
(\tilde{{\cal W}}_{\eta} + \tilde{{\cal W}}_K^0) -\frac{2}{3} q_c^2
(\tilde{{\cal W}}_{\omega}^T + \tilde{{\cal W}}_{K^*}^{T, \, 0})
\nonumber \\
P_{L, 1} & = & q^2 ({\cal W}_{\pi} + {\cal W}_K^1)
- (q^2 + \frac{1}{3} q_c^2)
(\tilde{{\cal W}}_{\pi} + \tilde{{\cal W}}_K^1) -\frac{2}{3} q_c^2
(\tilde{{\cal W}}_{\rho}^T + \tilde{{\cal W}}_{K^*}^{T, \, 1})
\nonumber \\
P_{T, 0} & = & q^2 ({\cal W}_{\omega}^T + {\cal W}_{K^*}^{T,\, 0})
- (q^2 + \frac{2}{3} q_c^2)
(\tilde{{\cal W}}_{\omega}^T + \tilde{{\cal W}}_{K^*}^{T, \, 0})
-\frac{1}{3} q_c^2
(\tilde{{\cal W}}_{\eta} + \tilde{{\cal W}}_K^0)
\nonumber \\
P_{T, 1} & = & q^2 ({\cal W}_{\rho}^T + {\cal W}_{K^*}^{T,\, 1})
- (q^2 + \frac{2}{3} q_c^2)
(\tilde{{\cal W}}_{\rho}^T + \tilde{{\cal W}}_{K^*}^{T, \, 1})
-\frac{1}{3} q_c^2
(\tilde{{\cal W}}_{\pi} + \tilde{{\cal W}}_K^1)
\nonumber \\
P_{C, 0} & = &
{\cal W}_{\omega}^C-\tilde{{\cal W}}_{\omega}^C
+{\cal W}_{K^*}^{C, \, 0}-\tilde{{\cal W}}_{K^*}^{C, \, 0}
\nonumber \\
P_{C, 1} & = &
{\cal W}_{\rho}^C-\tilde{{\cal W}}_{\rho}^C
+{\cal W}_{K^*}^{C, \, 1}-\tilde{{\cal W}}_{K^*}^{C, \, 1}
\nonumber \\
\end{eqnarray}
where the meaning of the tilde is,
\begin{equation}
\label{srcp2}
\tilde{{\cal W}}(q) = {\cal W}(\v{q}^2 \rightarrow
q_c^2 + \v{q}^2)
\end{equation}
the final expression for the interaction is given by
Eq.~(\ref{allmespot})


%

\newpage

Table I: Proton and neutron decay widths for $^{12}_\Lambda C$
in units of $\Gamma^0= 2.52 \cdot 10^{-6}$ eV.	The  direct contributions
are given in columns {\it dir.}, while the
results for the total transition rates (direct plus exchange),
obtained in the {\it Calculations I} and {\it II} (see text),
are listed  in columns	{\it Cal. I} and {\it Cal. II},  respectively.
\begin{center}
\begin{tabular}{cccccccccccc}	\hline\hline
meson & \multicolumn{3}{c}{$\Gamma_n$}& &
\multicolumn{3}{c}{$\Gamma_p$}& &
\multicolumn{3}{c}{$\Gamma_{n/p}$}
\\ \cline{2-4} \cline{6-8} \cline{10-12}
& {\it dir.} & {\it Cal. I} & {\it Cal. II} & & {\it dir.} &
{\it Cal. I} & {\it Cal. II} & & {\it dir.} & {\it Cal. I} &
{\it Cal. II} \\ \hline
$\pi$	   & 0.191 & 0.133 & 0.113 && 0.954 & 1.184 & 1.266 && 0.200 & 0.113 & 0.089 \\
$+ \eta$   & 0.240 & 0.160 & 0.110 && 0.924 & 1.119 & 1.152 && 0.260 & 0.143 & 0.095 \\
$+ K$	   & 0.192 & 0.255 & 0.256 && 0.648 & 0.657 & 0.734 && 0.295 & 0.389 & 0.349  \\
$+ \rho$   & 0.175 & 0.276 & 0.267 && 0.579 & 0.498 & 0.535 && 0.302 & 0.554 & 0.499   \\
$+ \omega$ & 0.315 & 0.373 & 0.332 && 0.683 & 0.606 & 0.594 && 0.461 & 0.616 & 0.559 \\
$+ K^*$    & 0.271 & 0.427 & 0.380 && 0.986 & 0.780 & 0.978 && 0.274 & 0.547 & 0.389 \\
\hline\hline \\
\end{tabular}

\end{center}

\newpage

Table II: Contribution of the $\rho$-meson in the nonmesonic decay
of $^{12}_\Lambda C$.  $\Gamma^{PC}$ and $\Gamma^{PV}$
stand for the
parity conserving  and parity violating  rates, respectively.
Units are the same as in Table I.
\begin{center}

\begin{tabular}{cccccc}
       \hline\hline
\multicolumn{6}{l} {$\pi$}   \\ \hline
~~~~~~~~~~~~~~~  & ~~~~~$\Gamma_n^{PC}$~~~~~ &
~~~~~$\Gamma_n^{PV}$~~~~~ & ~~~~~$\Gamma_p^{PC}$~~~~~
& ~~~~~$\Gamma_p^{PV}$~~~~~ & ~~~~~$\Gamma_\Lambda$~~~~~  \\ \hline
Ref. \cite{ba02}  & 0.009  & 0.151 & 0.734   & 0.383  & 1.277  \\
{\it Cal. II}& 0.005  & 0.108 & 1.004	& 0.262  & 1.379  \\   \hline \\
\multicolumn{6}{l} {$\rho$}   \\ \hline
Ref. \cite{ba02}  & 0.005  & 0.003 & 0.109   & 0.008  & 0.125  \\
{\it Cal. II}& 0.007  & 0.003 & 0.100	& 0.012  & 0.122  \\   \hline \\
\multicolumn{6}{l} {$\pi + \rho$}   \\ \hline
Ref. \cite{ba02}  & 0.009  & 0.133 & 0.583   & 0.461  & 1.186  \\
{\it Cal. II}& 0.004  & 0.128 & 0.727	& 0.204  & 1.063  \\  \hline \hline \\
\end{tabular}

\end{center}

\newpage

Table III: Contribution  of the $K$-meson.  The cutoffs in Ref. \cite{sa00} are:
$\Lambda_{\pi}=\Lambda_{K}=1.300$ MeV. Units are the same as in Table I.

\begin{center}

\begin{tabular}{cccccc}       \hline\hline
\multicolumn{4}{l} {~~~~~$\pi$}   \\ \hline
~~~~~~~~~~~~~~~~~~~~  & ~~~~~~~~$\Gamma_n$~~~~~~~~ &
~~~~~~~~ $\Gamma_p$~~~~~~~~ & ~~~~~~~$\Gamma_n/\Gamma_p$ ~~~~~~~\\ \hline
Ref. \cite{sa00}  & 0.221 & 2.354 & 0.094  \\
Ref. \cite{ji01}    & 0.119 & 0.956 & 0.124  \\
{\it Cal. I} & 0.133 & 1.184 & 0.113  \\
{\it Cal. I'}        & 0.120 & 1.090 & 0.110  \\
{\it Cal. II}	 & 0.113 & 1.266 & 0.089  \\  \hline \\
\multicolumn{4}{l} {~~~~~$\pi + K$}  \\ \hline
Ref. \cite{sa00}  & 0.459 & 1.300 & 0.353  \\
Ref. \cite{ji01}    & 0.273 & 0.522 & 0.523  \\
{\it Cal. I} & 0.217 & 0.697 & 0.311  \\
{\it Cal. I'}        & 0.223 & 0.485 & 0.460  \\
{\it Cal. II}	 & 0.229 & 0.802 & 0.285  \\  \hline \hline
 \\
\end{tabular}

\end{center}

\newpage

Table IV: Contributions of individual mesons to the decay width
for $^{12}_\Lambda C$.	Units are
the same as in Table I.

\begin{center}

\begin{tabular}{cccccc}       \hline\hline
meson &\multicolumn{2}{c}{$\Gamma_n$} &
&\multicolumn{2}{c}{$\Gamma_p$}
\\ \cline{2-3} \cline{5-6}
 & Ref.\cite{ba02} & {\it Cal. II} &
& Ref.\cite{ba02} & {\it Cal. II}    \\ \hline
$\pi$ & 0.159 & 0.113 && 1.107 & 1.266 \\
$\eta$ & 0.007 & 0.004 && 0.009 & 0.009 \\
$K$ & 0.076 & 0.048 && 0.139 & 0.157 \\
$\rho$ & 0.008 & 0.010 && 0.116 & 0.112 \\
$\omega$ & 0.011 & 0.069 && 0.069 & 0.150 \\
$K^*$ & 0.058 & 0.168 && 0.083 & 0.268 \\ \hline
$\pi+\eta$ & 0.215 & 0.110 && 1.004 & 1.152 \\
$\pi+K$ & 0.269 & 0.229 && 0.830 & 0.802 \\
$\pi+\rho$ & 0.141  & 0.132 && 1.035 & 0.932 \\
$\pi+\omega$ & 0.189 & 0.174 && 1.308 & 1.465 \\
$\pi+K^*$ & 0.118 & 0.359 && 1.462 & 2.050 \\ \hline $all \;
mesons$ & 0.275 & 0.380 && 1.061 & 0.978 \\ \hline \hline
\end{tabular}

\end{center}

\newpage
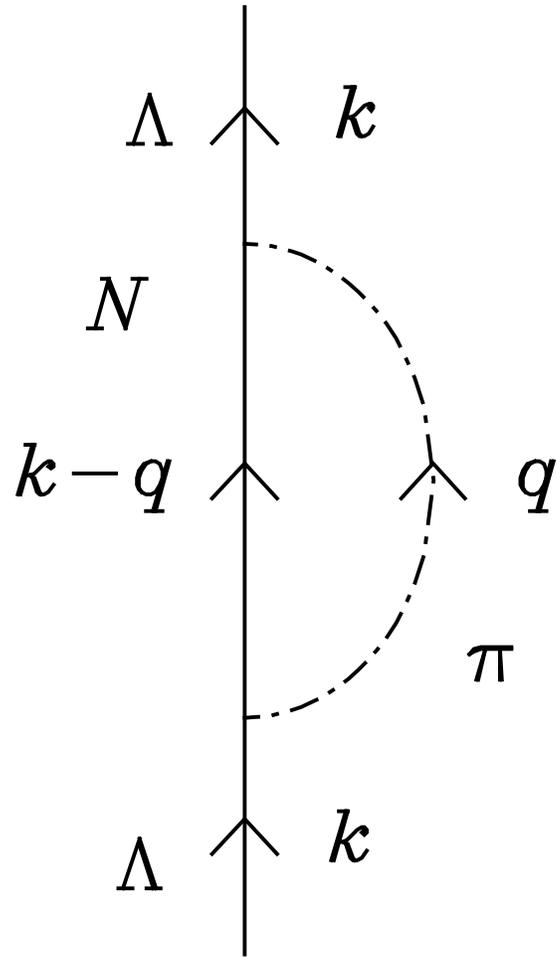
\begin{figure}
\caption{$\Lambda$ self-energy in nuclear matter.
Dot-dashed line represents a dressed-pion in nuclear
matter. The continuous lines stand either for a nucleon or
for the $\Lambda$ (as indicated in the figure).}
\label{fig:self1}
\end{figure}
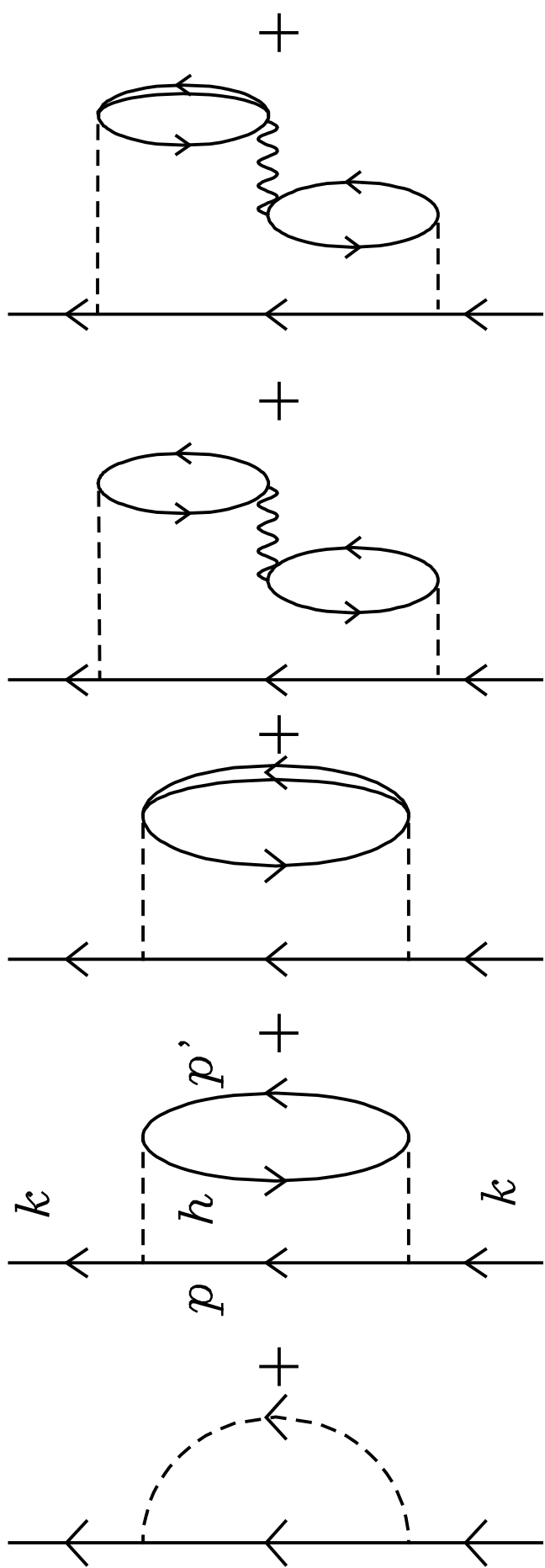
\begin{figure}
\caption{A few lowest order terms for the
$\Lambda$ self-energy in nuclear matter. The dotted and wavy
lines represent, respectively, the  undressed  pion and  $NN$ strong interaction,
while the $\Delta$ excitation is denoted by the double continuous line.}
\label{fig:self2}
\end{figure}
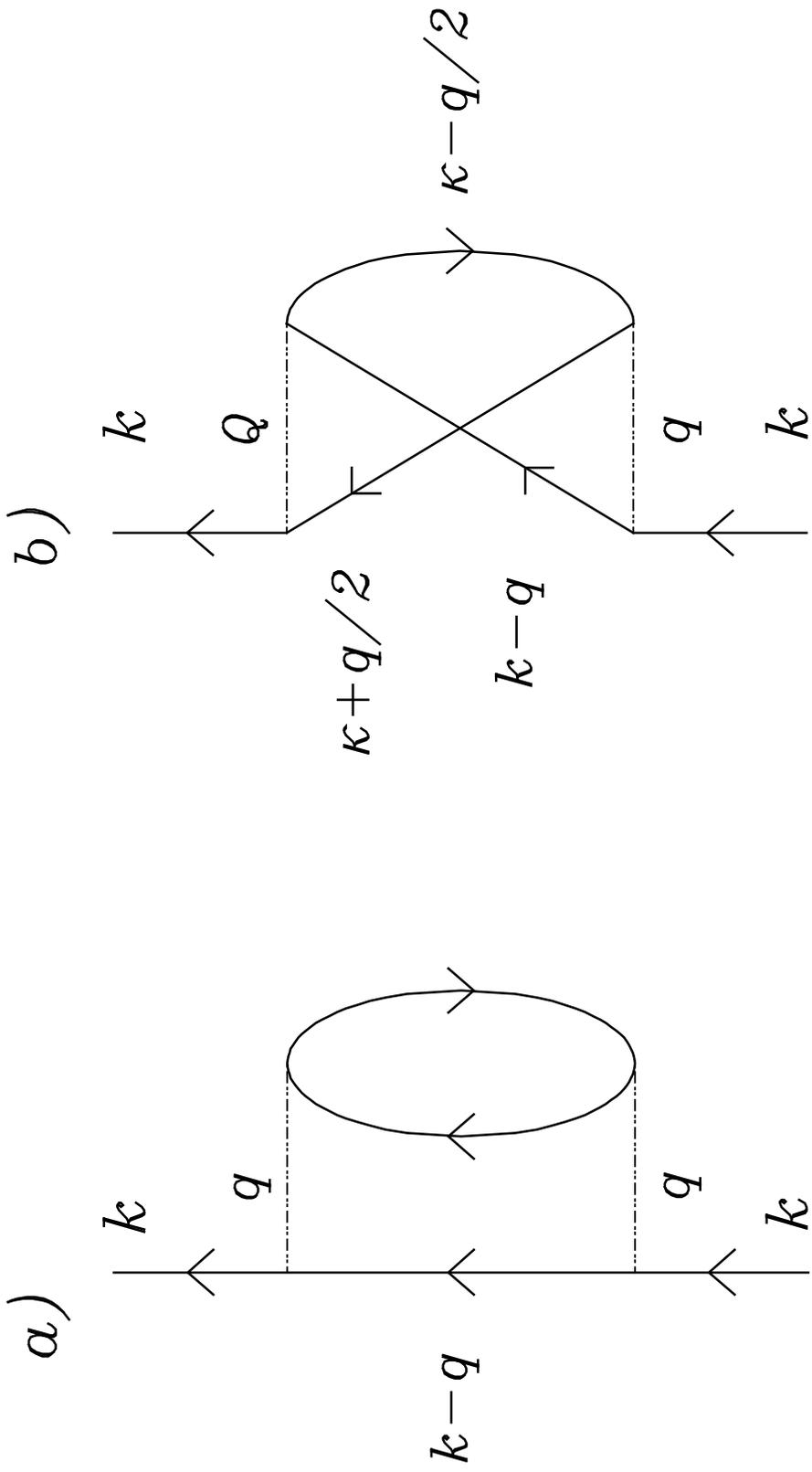
\begin{figure}
\caption{Direct ($a$) and exchange ($b$) contributions
to the $\Lambda$ decay width. The dashed-double dotted lines represent
the full $(\pi + \eta + K + \rho + \omega + K^*)$-transition
potential.}
\label{fig:direxc}
\end{figure}

\end{document}